\documentclass{emulateapj}
\slugcomment{Accepted: April 28, 2016}
\shorttitle{Weibel instability driven by spatially anisotropic density structures}

\shortauthors{Sara Tomita and Yutaka Ohira}

\begin{document}

\title{Weibel instability driven by spatially anisotropic density structures}

\author{Sara Tomita and Yutaka Ohira}

\begin{abstract}
Observations of afterglows of gamma-ray bursts suggest (GRBs) that post-shock magnetic fields are strongly amplified to about 100 times the shock-compressed value. 
The Weibel instability appears to play an important role in generating of the magnetic field. 
However, recent simulations of collisionless shocks in homogeneous plasmas show that the magnetic field generated by the Weibel instability rapidly decays. 
There must be some density fluctuations in interstellar and circumstellar media. 
The density fluctuations are anisotropically compressed in the downstream region of relativistic shocks. 
In this paper, we study the Weibel instability in electron--positron plasmas with the spatially anisotropic density distributions by means of two-dimensional particle-in-cell simulations. 
We find that large magnetic fields are maintained for a longer time by the Weibel instability driven by the spatially anisotropic density structure. 
Particles anisotropically escape from the high density region, so that the temperature anisotropy is generated and the Weibel instability becomes unstable. 
Our simulation results suggest that the Weibel instability driven by an anisotropic density structure can generate sufficiently large magnetic fields and they can cover sufficiently large regions to explain the afterglow emission of GRBs. 
\end{abstract}

\keywords{instabilities ---
magnetic fields ---
plasmas ---
shock waves ---
gamma-ray burst: general}

\affil{Department of Physics and Mathematics, Aoyama Gakuin University, 
5-10-1 Fuchinobe, Sagamihara 252-5258, Japan; tomisara@phys.aoyama.ac.jp}
\section{INTRODUCTION}
\label{sec:1}
The Weibel instability generates magnetic fields in collisionless plasmas with the temperature anisotropy \citep{weibel59}, 
which is thought to be important for the generation of magnetic fields and the acceleration of particles in relativistic shocks \citep{medvedev99,spitkovsky08b}. 
Furthermore, the Weibel instability is expected to have an important role even in nonrelativistic shocks \citep{kato10,matsumoto15} and shocks produced in leaser experiments \citep{fox03,huntington15}. 

A recent analysis of the afterglows of gamma-ray bursts (GRBs) shows that 
the ratio of the magnetic field energy density to the shock kinetic energy density, $\epsilon_{\rm B}$, 
has a range of about $10^{-8}-10^{-3}$ and median of about $10^{-5}$ \citep{santana14}. 
This means that magnetic fields in the downstream region of external forward shocks are, on average, 
amplified to 100 times the shock-compressed value. 
Although the $\epsilon_{\rm B}$ of GRB afterglows may not be estimated accurately, 
many GRB afterglows suggest the magnetic field amplification as long as we apply the standard model of GRB afterglows \citep{sari98}.
The time scale of the afterglow emission (the deceleration time of blastwaves) in the downstream rest frame is given by \citep{sironi07}
\begin{equation}
t_{\rm dec} \omega_{\rm pp} = 7.4 \times 10^7~E_{\rm iso,53}^{1/3} n_0^{1/6} \Gamma_{2}^{-5/3}
\end{equation}
where $\omega_{\rm pp}$ is the plasma frequency of protons, $E_{\rm iso}=10^{53}E_{\rm iso,53}~{\rm ergs}$ is the isotropic equivalent energy of the GRB, $n=10^0n_{0}~{\rm cm}^{-3}$ is the upstream number density in the upstream rest frame, and $\Gamma_2=10^2\Gamma$ is the Lorentz factor of the blastwave. 
However, recent particle-in-cell (PIC) simulations of relativistic shocks in homogeneous electron--positron plasmas showed that the magnetic field produced by the Weibel instability cannot occupy large downstream regions because of rapid decay \citep{chang08}. 
PIC simulations by \citet{chang08} showed that the time evolution is described by $\epsilon_{\rm B}\propto (t\omega_{\rm pe})^{-1}$, where $\omega_{\rm pe}$ is the plasma frequency of the electrons, so that $\epsilon_{\rm B}$ becomes much smaller than that required by observations at $t=t_{\rm dec}$. 
Theoretical studies about the decay showed that the decay rate strongly depends on the power spectrum of magnetic field fluctuations \citep{chang08,lemoine15}. 
According to \citet{chang08}, it is given by $\epsilon_{\rm B} \propto (t\omega_{\rm pe})^{-2(p+1)/3}$, 
where $p$ is the spectral index of magnetic field fluctuations ($|\delta B_k|^2 \propto k^p$). 
Interestingly, \citet{keshet09} showed using long PIC simulations that the decay of magnetic fields generated by the Weibel instability becomes slower once particles are accelerated by the Weibel mediated shock. 
The simulation time of \citet{keshet09} is about $10^4~\omega_{\rm pe}^{-1}$, which is much smaller than the time scale of GRB afterglows.
Therefore, whether the magnetic field generated by the Weibel instability can explain observations of afterglows is an open question. 
In addition, why the distribution of $\epsilon_{\rm B}$ spans five orders of magnitude is also an interesting question. 
If $\epsilon_{\rm B}$ was controlled only by the microphysics of a uniform plasma, 
$\epsilon_{\rm B}$ would have a characteristic value. 
Since GRBs are expected to be produced in supernova-like environments \citep{anderson15}, 
there may be some macroscopic structures in GRB environments, for example, clumpy wind regions, 
shocked wind regions, and so on. 
Furthermore, as with supernovae, each GRB may have very different length scales of these macroscopic structures. 
Hence, the wide distribution of $\epsilon_{\rm B}$ suggests that in addition to microphysics, 
the macroscopic properties of GRB environments are important for the generation of magnetic fields.

The turbulent dynamo in the shock downstream region is proposed as another magnetic field amplification mechanism  \citep{sironi07,inoue11,mizuno11}. 
There must be some density fluctuations in the circumstellar (CSM) and interstellar media (ISM).  
The injection scale of turbulence is about $100~{\rm pc}$ for the ISM \citep{armstrong95}.
For the CSM, it must be smaller than that of the ISM but the typical scale is not clear \citep[e.g. see section 3 of ][]{sironi07}. 
External forward shocks interact with the density fluctuations, so that turbulence is driven and magnetic fields are amplified by the turbulence. 
Relativistic magnetohydrodynamics simulations showed that the growth and decay time scales of magnetic field fluctuations are comparable to the time in which plasma crosses the length scale of the density fluctuations with the turbulent velocity (the so-called eddy turnover time) \citep{inoue11}. 
This is because the eddy motion stretches magnetic field lines and the decay time scale of the turbulence is also the eddy turnover time.

In the downstream region of relativistic shocks, 
the isotropic density fluctuations in the upstream region are strongly compressed in the shock-normal direction, 
so that the compressed density structures have large spatial anisotropy. 
Then, particles in the high density region anisotropically escape to the shock-normal direction (see Section~\ref{sec:2} for details). 
As a result, a temperature anisotropy is generated and the Weibel instability is expected to become unstable in the high and low density regions. 
Therefore, in addition to the shock transition region, the Weibel instability is expected to generate magnetic fields in the downstream region of shocks propagating into inhomogeneous media.  
In this paper, by using two-dimensional PIC simulations, we study the nonlinear evolution of the Weibel instability in spatially anisotropic density structures (Section \ref{sec:3}). 
Our results show that the Weibel instability driven by the spatially anisotropic density structure generates magnetic fields and suggest that the generated magnetic fields cover large regions where GRB afterglows are emitted (Section \ref{sec:4}). 

\section{GENERATION OF THE TEMPERATURE ANISOTROPY}
\label{sec:2}
\begin{figure*}
\plotone{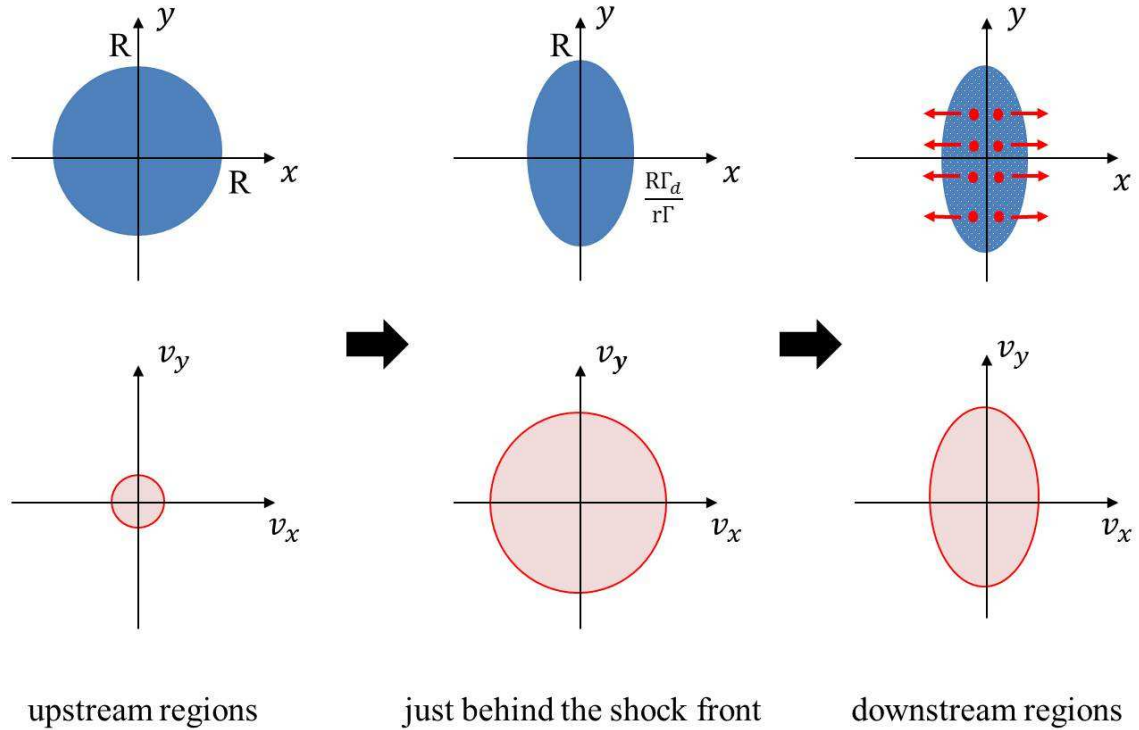} 
\caption{Generation mechanism of the temperature anisotropy in the shock downstream region. 
The upper and lower panels show the structure of a high density region and the velocity distribution in the high density region, respectively. The left, middle, and right columns show the shock upstream region, the region just behind the shock, and the downstream region, respectively. The cartoons are all in the local fluid rest frame. 
\label{fig:1}}
\end{figure*}

In this section, we explain how the temperature anisotropy is generated in the shock downstream region. 
Figure~\ref{fig:1} shows cartoons illustrating the structure of a high density clump and the velocity distribution in the high density clump. 
All the cartoons are illustrated in the local rest frame and the shock-normal direction is the $x$-direction. 
We consider the evolution of an isotropic high density clump in the upstream region where the velocity distribution is isotropic and cold (the left column). 
After the high density clump passes through the shock front (the middle column), the high density clump is strongly compressed by $\Gamma_{\rm d} / (\Gamma r)$ in the $x$ direction at the downstream rest frame, where $r$ and $\Gamma_{\rm d}$ are the shock compression ratio and the Lorentz factor of the downstream flow at the shock rest frame. 
For unmagnetized shocks, the Weibel instability generates magnetic field fluctuations and dissipates the upstream bulk flow in the shock transition region, so that the upstream plasma is heated almost isotropically. 
Then, the downstream plasma has a large velocity dispersion, so that particles start to escape from the high density clump in the shock downstream region (the right column). 
Since the high density clump is contracted by $\Gamma_{\rm d} / (\Gamma r)$ in the shock-normal direction ($x$ direction), 
particles with a large $v_{\rm x}$ initially escape from the high density clump in the $x$ direction but it takes longer time to escape in the shock-tangential direction. 
As a result, a temperature anisotropy appears at the high density clump in the downstream region. 
In the high density clump, the temperature in the shock-normal direction becomes lower than that in the shock-tangential direction. 
On the other hand, outside the high density clump, the temperature in the shock-normal direction becomes higher than that in the shock-tangential direction because there are escaping particles with a large $v_{\rm x}$. 
Hence, the upstream density fluctuations generate the temperature anisotropy in the shock downstream region.

The generation time scale of the temperature anisotropy is the sound crossing time, 
\begin{equation}
t_{\rm ta}= \frac{R\Gamma_{\rm d}}{\Gamma r v_{\rm th}}~~, 
\end{equation}
where $v_{\rm th}$ and $R$ are the thermal velocity in the downstream region and the size of the high density clump in the upstream region, respectively. 
The temperature anisotropy drives the Weibel instability and generates a magnetic field. 
To explain the afterglow emission of GRBs, the characteristic time scale, $t_{\rm ta}$, should be comparable to $t_{\rm dec}$, that is, 
the length scale of density fluctuations should be 
\begin{equation}
R\approx  4.8 \times 10^{17}~{\rm cm}~\left(\frac{r/\Gamma_{\rm d}}{2\sqrt{2}}\right) \frac{v_{\rm th}}{c} E_{\rm iso,53}^{1/3} n_0^{-1/3} \Gamma_2^{-2/3}~~,
\end{equation}
where $r/\Gamma_{\rm d}=2\sqrt{2}$ and $4$ for relativistic and nonrelativistic strong shock limits, respectively. 

Although the injection scale of the ISM turbulence is much larger than the above estimation, the ISM turbulence has the Kolmogorov like spectrum \citep{armstrong95}. 
So that there are density fluctuations with a smaller scale in the ISM. 
In addition, the CSM is thought to have density fluctuations with a scale smaller than that of ISM. 
Moreover, if particles are accelerated by shocks, the accelerated particles would generate the density fluctuations \citep{niemiec08,ohira09,riquelme09,ohira13,ohira14,ohira16}. 
However, the characteristic length scale is still uncertain because the nonlinear evolution is an open question, but one of expected length scales is the gyroradius of the accelerated particle that depends on the particle energy and magnetic fields.   
In any case, we can expect many types of the density fluctuations although the characteristic length scale is still uncertain. 
Therefore,  in addition to the narrow shock transition region, we expect that the Weibel instability generates magnetic fields in the wide downstream region if the shock propagates to inhomogeneous plasmas. 
\section{SIMULATION}
\label{sec:3}
\subsection{Setting}
We use the two-dimensional electromagnetic PIC code, pCANS, 
in order to study the nonlinear evolution of the Weibel instability in inhomogeneous plasmas. 
The pCANS use the density decomposition method for the calculation of the current density. 
We use the first order weighting method (the could in cell method) in this simulation.
We set a two-dimensional simulation box in the $xy$-plane with the periodic boundary condition in both directions. 
We track three components of particle velocity, magnetic fields, and electric fields.

Initially, we prepare two counter-streaming unmagnetized electron--positron plasmas with the drift velocities, $v_{\rm d}= \pm 0.5 c$, where the counter-streaming direction is the $y$-direction and $c$ is the speed of light. 
Each electron--positron plasma has the same density and thermal velocity of $v_{\rm th} = 0.1c$. 
This initial velocity distribution is unstable for the Weibel instability. 
The shock velocity is relativistic for the early phase of GRB afterglows and decelerated to the nonrelativistic velocity as time goes on. 
The relativistic flow in PIC simulations generates the numerical Cherenkov radiation \citep{godfrey74}. 
To avoid worry about how the numerical Cherenkov radiation affects 
the long time evolution of our simulations, we chose a mildly relativistic velocity, $0.5c$.
We believe that whether this is relativistic or nonrelativistic is unimportant 
because the important point is the generation of the temperature anisotropy.
Previous PIC simulations showed that the electron temperature becomes almost the same as that of ions in the downstream region of relativistic shocks \citep{spitkovsky08a,kumar15}. 
In addition, the electron--positron plasma is simple and easy to simulate compared with 
the electron-proton plasma. 
Therefore, we consider here the electron--positron plasma as a first step.

Since the surroundings of GRB progenitors have not been understood well, 
we assume the simplest density structure to describe the downstream density distribution. 
We consider two density structures, $n(x,y)$, in this paper:
\begin{enumerate}
\item $n(x,y)=n_0$
\item $n(x,y)=n_0 \left\{1+0.5\sin(2\pi x/L_{\rm x})\right\}$
\end{enumerate}
One is the homogeneous distribution, where $n_0$ is the constant density. 
The other is the spatially anisotropic density distribution depending only on the $x$-coordinate, 
where $L_{\rm x}$ is the simulation box size in the $x$-direction 
(i.e. there is one wavelength in the simulation box) 
and the amplitude of the density fluctuation is $\delta n / n_0=0.5$.  
The length scale of the high density region is $L_{\rm x}/2$ which corresponds to $R \Gamma_d / ( \Gamma r )$ in Section~\ref{sec:2}. 
We perform three runs for spatially anisotropic density structures 
to investigate the length dependence of the magnetic field generation, 
$L_{\rm x}=L_{\rm y}=120, 240,$ and $480~c/\omega_{\rm pe}$, 
where $\omega_{\rm pe}$ is the plasma frequency defined by the mean number density of electron--positron plasmas, $n_0$. 
The cell size and time step of simulations are $\Delta x = \Delta y = 0.1~c / \omega_{\rm pe}$ and $\Delta t = 0.05~\omega_{\rm pe}^{-1}$, respectively. 
Initially, each cell contains 40 electrons and 40 positrons on average. 

\subsection{Results}
\begin{figure}
\plotone{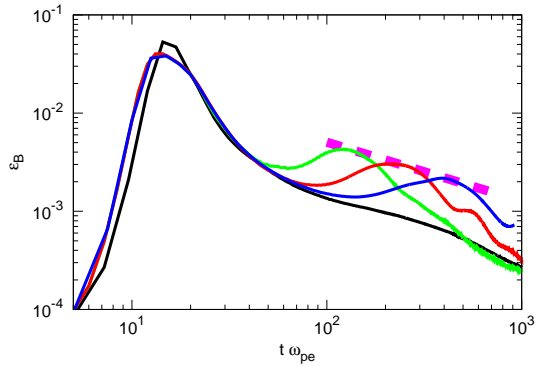} 
\caption{Time evolution of the mean magnetic field energy density normalized by the mean initial kinetic energy density, $\epsilon_{\rm B}$. 
The black line shows $\epsilon_{\rm B}$ for the homogeneous density distribution. 
The green, red, and blue lines show $\epsilon_{\rm B}$ for the anisotropic density distributions with $L_{\rm x}=120, 240 , 480~c/\omega_{\rm pe}$, respectively. 
The magenta dashed line shows the envelope of the second peaks of the three simulations, $\epsilon_{\rm B}=0.08(t\omega_{\rm pe})^{-0.6}$. 
\label{fig:2}}
\end{figure}

In Figure~\ref{fig:2}, we show the time evolution of the mean energy density of magnetic fields normalized 
by the mean initial kinetic energy density, $\epsilon_{\rm B}$. 
The initial condition of the counter-streaming flow is unstable for the Weibel instability. 
For the homogeneous density distribution (black line), initially, 
the magnetic field grows exponentially as predicted by the linear analysis of the Weibel instability. 
After the saturation at $t \sim10~\omega_{\rm pe}^{-1}$, the magnetic field simply decays and the velocity dispersion becomes large. 
It should be noted that the long time evolution of $\epsilon_{\rm B}$ in this simulation is not reliable because of the numerical noise and the finite size of simulation box. 
The noise level is $\epsilon_B \sim 5\times 10^{-4}$ in this simulation. 

On the other hand, for the anisotropic density structures with $L_{\rm x}=120$ (green line), 
240 (red line), and $480~c/\omega_{\rm pe}$ (blue line), 
after the first saturation at $t\sim10~\omega_{\rm pe}^{-1}$, 
the magnetic field starts to grow again from $t\sim 10^2~\omega_{\rm pe}^{-1}$, and then 
the magnetic field decays after the second saturation. 
The time scale of the second generation of magnetic fields is comparable to the sound crossing time of the high density structure, $L_{\rm x}/2v_{\rm th}$, and becomes longer as the length scale of the density structure becomes larger. 
These are consistent with our expectation in Section~\ref{sec:2}. 
Therefore, interestingly, the magnetic field is generated for a longer time in the anisotropic density distribution 
even though the initial kinetic energy is the same as that of the homogeneous distribution. 
Hereafter, we show simulation results for the anisotropic density distribution with $L_{\rm x}=240~c/\omega_{\rm pe}$.

\begin{figure}
\plotone{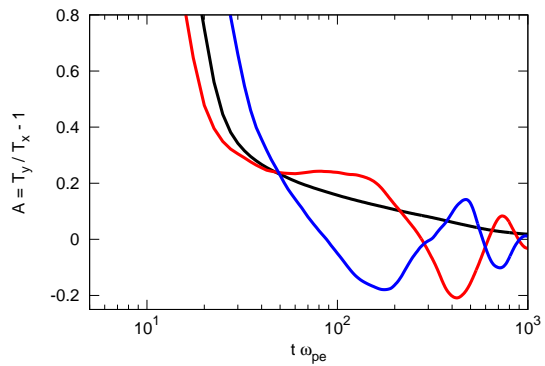} 
\caption{Time evolution of the temperature anisotropy, $A=T_{\rm y}/T_{\rm x}-1$, for the anisotropic density distribution with $L_{\rm x}=240~c/\omega_{\rm pe}$. 
The red and blue lines show $A$ in the high ($50~c/\omega_{\rm pe}<x<70~c/\omega_{\rm pe}$) and low  ($170~c/\omega_{\rm pe}<x<190~c/\omega_{\rm pe}$) density regions, respectively.  
The black line shows A in the whole simulation box ($0<x<240~c/\omega_{\rm pe}$) for the homogeneous density distribution. 
\label{fig:3}}
\end{figure}

In order to understand the origin of the second growth of the magnetic field, in Figure~\ref{fig:3}, 
we show the time evolution of the temperature anisotropy, $A=T_{\rm y}/T_{\rm x}-1$, in the high (red line) and low (blue line) density regions. 
For the homogeneous density distribution (black line), the temperature anisotropy decreases with time monotonically and closes to $0$. 
On the other hand, in the high density region for the anisotropic density distribution (red line), 
the temperature anisotropy becomes $A\sim 0.25$ at $t\sim 10^2~\omega_{\rm pe}^{-1}$, which is larger than that of the homogeneous density distribution. 
This is because particles moving to the $x$-direction escape to the low density region, 
0.0.so that the temperature in the $x$-direction decreases. 
Furthermore, the temperature in the $x$-direction increases with time in the low density region (blue line) at $t\sim 10^2~\omega_{\rm pe}^{-1}$ because the particles moving to the $x$ direction accumulate in the low density region. 
After $t\sim 10^2~\omega_{\rm pe}^{-1}$, the temperature anisotropy oscillates with the sound crossing time as the amplitude decreases. 

\begin{figure}
\plotone{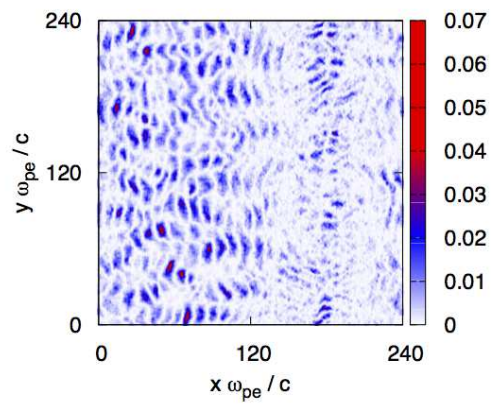} 
\caption{Spatial distribution of the magnetic energy density normalized by the mean initial kinetic energy density, $\epsilon_{\rm B}$, for the anisotropic density distribution with $L_{\rm x}=240~c/\omega_{\rm pe}$ at $t=200~\omega_{\rm pe}^{-1}$. 
\label{fig:4}}
\end{figure}

In Figure~\ref{fig:4}, we show the spatial distribution of $\epsilon_{\rm B}$ at the time of the second saturation, 
$t=200~\omega_{\rm pe}^{-1}$. 
The magnetic fields are efficiently generated in the high ($x\sim L_{\rm x}/4$) 
and low ($x\sim3L_{\rm x}/4$) density regions. 
The wave vector is parallel to the $x$-direction in the high density region, 
while it is parallel to the $y$-direction in the low density region. 
The wavelength of the magnetic field fluctuation is typically about $20~ c/\omega_{\rm pe}$. 

For a small temperature anisotropy ($A\ll1$), the maximum growth rate of the Weibel instability is given by \citep{davidson72}
\begin{equation}
\gamma_{\rm max}=\left(\frac{4}{27\pi}\right)^{1/2}\frac{A^{3/2}}{1+A}\frac{v_{\rm th}}{c}\omega_{\rm pe} 
\end{equation}
for $k = (A/3)^{1/2}~\omega_{\rm pe}/c$, where $k$ is the wave vector of the most unstable mode and parallel to the direction of the lower temperature. 
For $A= 0.25$ and $v_{\rm th} = 0.5~c$, the mode with $k \approx 0.3~\omega_{\rm pe}/c$ has the maximum growth rate of $\gamma_{\rm max}\approx 10^{-2}~\omega_{\rm pe}$, which are consistent with our simulation results. 
Therefore, we conclude that the second generation of magnetic fields in the anisotropic density distribution is due to the Weibel instability and the temperature anisotropy after the first saturation is produced by the anisotropic escape of particles. 

\section{DISCUSSION}
\label{sec:4}
In Section~\ref{sec:3}, we have showed that the magnetic field is generated for a long time in the anisotropic density distribution compared with the homogeneous density distribution. 
In order to compare with observed properties of GRB afterglows, 
we have to perform a much larger simulation, 
$L_{\rm x}\sim 10^7~c/\omega_{\rm pp}$ and $t\sim 10^7~\omega_{\rm pp}^{-1}$. 
Since the larger simulation is very challenging for current supercomputers, 
we here assume a scaling law obtained by fitting our limiting results of PIC simulations. 
The dashed line in Figure~\ref{fig:2} shows the scaling law that is the envelope of the second peaks of the three simulations 
and it is given by
\begin{equation}
\epsilon_{\rm B} \approx 0.08 (t\omega_{\rm pe})^{-0.6}~~.
\end{equation}
If there are anisotropic density distributions with the length scale of $10^7 c/\omega_{\rm pp}$ in the downstream region, 
the above scaling suggests that magnetic fields are generated up to $\epsilon_{\rm B}\sim 10^{-5}$ for the time scale of $t\sim10^7~\omega_{\rm pp}^{-1}$. 
As mentioned in Section~\ref{sec:2}, there would be many types of inhomogeneity, for example, inhomogeneity in stellar winds and the ISM. 
Moreover, particles accelerated by shocks would generate inhomogeneity. 
A variety of environments of GRB progenitors could distribute $\epsilon_{\rm B}$ in the wide range. 
These are consistent with observations of GRB afterglows. 
Therefore, the post-shock magnetic fields of GRB afterglows could arise from the Weibel instability driven by shock-compressed anisotropic density distributions. 
In order to confirm the above scaling, we need larger simulations. 
In addition, we have to confirm our results by performing simulations of collisionless shocks propagating into inhomogeneous media. 

If there are the density fluctuations in the upstream region, 
the magnetic field is amplified by the turbulent dynamo. 
The main difference between the turbulent dynamo and our model is the coherent length scale of amplified fields. 
Our model predicts that the length scale is of the order of $10~c/\omega_{\rm pp}$
,which is much smaller than the coherent length scale of the turbulent dynamo. 
The difference should be reflected in the afterglow emission because 
the emission mechanism is not the synchrotron radiation but the jitter 
radiation \citep{medvedev00} in our model. 
Furthermore, the growth time scale of the turbulent dynamo is about the eddy turnover time that is slightly longer than that of our model (the sound crossing time) because the downstream turbulence is subsonic.  
Therefore, nonlinear interactions between the small scale turbulence generated by the Weibel instability and the large scale turbulence are also an interesting problem. 
We plan to study relativistic collisionless shocks propagating into inhomogeneous media in the future. 

In this paper, we used electron--positron plasmas although electron--ion plasmas are realistic for many situations. 
As long as the shock velocity is relativistic enough, electron--ion plasmas can be treated as electron--positron plasmas in the downstream region\citep{spitkovsky08a,kumar15}. 
However, at the late phase of GRB afterglows, the shock velocity is not so relativistic and the difference between electron--positron plasma and electron--proton plasma would be significant \citep[e.g.][]{stockem15}.

On the other hand, the electron heating in nonrelativistic shocks is an open question 
\citep{ghavamian07,ohira07,ohira08,rakowski08}. 
Furthermore, the background magnetic field becomes more important for the nonrelativistic shocks, 
which would suppress the escape of particles from the high density region and 
stabilize the Weibel instability, but the whistler, mirror, and Alfv{\'e}n ion cyclotron instabilities would be unstable. 
In any case, the anisotropic density structure would make some kinetic instabilities unstable and excite some plasma waves in any plasmas because the temperature anisotropy is generated by the anisotropic escape \citep[e.g.][]{rosenbluth56,chandrasekhar58,vedenov58}. 
Since plasma waves play an important role in the heating of plasmas, and the acceleration and propagation of cosmic rays, 
the kinetic instabilities driven by the anisotropic density structure could be important for cosmic-ray physics. 
 
\section{SUMMARY}
\label{sec:5}
In this paper, we have showed that anisotropic density structures in collisionless plasmas 
produce a temperature anisotropy, such that magnetic fields are generated by the Weibel instability. 
Such a situation can be expected in the downstream region of relativistic shocks propagating into inhomogeneous plasmas 
because the isotropic inhomogeneity in the upstream region is anisotropically compressed by the shocks. 
Thanks to this mechanism, the magnetic fields generated by the Weibel instability cover larger 
downstream regions than previously thought. 
Hence, the Weibel instability that we observed in this paper could be important for radiation from GRBs \citep{medvedev99}, particle accelerations in the Weibel mediated shocks \citep{spitkovsky08b} and generation of cosmological magnetic fields \cite{schlickeiser03,fujita05,medvedev06}. 

\acknowledgments
We thank the anonymous referee for fast and constructive comments, and R.~Yamazaki, J.~Shimoda, Y.~Shoji, M.~Nogami, and E.~Kobayashi for useful comments. 
We also thank ISSI (Bern) for support of the team "Physics of the injection of particle acceleration at  astrophysical, heliospheric, and laboratory collisionless shocks". 
The software used in this work was in part developed in pCANS at Chiba University. 
Numerical computations were carried out on Cray XC30 at Center for Computational Astrophysics(CfCA), National Astronomical Observatory of Japan. 
ST is supported by the Komoda fellowship. The page charges of this paper are supported by CfCA of NAOJ.

\end{document}